\documentclass{iau}
\usepackage{graphicx}

\title[Late-time evolution of ultracompact X-ray binaries] 
{Late-time evolution of ultracompact X-ray binaries}

\author[L.~M.~van~Haaften et al.]   
{L.~M.~van~Haaften$^{1,*}$,
 G.~Nelemans$^{1,2}$
 \and R.~Voss$^1$}

\affiliation{$^1$Department of Astrophysics/ IMAPP, Radboud University Nijmegen, P.O. Box 9010, 6500 GL Nijmegen, The Netherlands, \\ * email: {\tt L.vanHaaften@astro.ru.nl} \\[\affilskip]
$^2$Institute for Astronomy, KU Leuven, Celestijnenlaan 200D, 3001 Leuven, Belgium \\[\affilskip]}

\pubyear{2012}
\volume{290}  
\jname{Feeding compact objects: Accretion on all scales}
\editors{C.M. Zhang, T. Belloni, M. M\'endez \& S.N. Zhang, eds.}

\begin{document}

\maketitle

\begin{abstract}
Ultracompact X-ray binaries (UCXBs) have orbital periods shorter than about 80 minutes and typically consist of a neutron star that accretes hydrogen-poor matter from a white dwarf companion. Angular momentum loss via gravitational wave radiation drives mass transfer via Roche-lobe overflow. The late-time evolution of UCXBs is poorly understood -- all 13 known systems are relatively young and it is not clear why. One question is whether old UCXBs actually still exist, or have they become disrupted at some point? Alternatively they may be simply too faint to see. To investigate this, we apply the theories of dynamical instability, the magnetic propeller effect, and evaporation of the donor, to the UCXB evolution. We find that both the propeller effect and evaporation are promising explanations for the absence of observed long-period UCXBs.
\end{abstract}

\firstsection 
\section{Introduction}

Ultracompact X-ray binaries (UCXBs) are a subclass of low-mass X-ray binaries and usually consist of a white dwarf transferring mass to a neutron star or black hole companion \cite[(Savonije et al. 1986)]{savonije1986}. Today, about 13 systems are known, all Galactic. Apart from their short periods they distinguish themselves from other X-ray binaries by a very low optical to X-ray flux ratio.
UCXBs are important objects to study because of the absence of hydrogen in the accretion disk and X-ray bursts. They are also excellent test objects for the common-envelope phase, which usually happens twice in the process of their formation. UCXBs are also likely progenitors of millisecond radio pulsars.

During their evolution, UCXBs transfer mass via Roche-lobe overflow because they lose angular momentum via the emission of gravitational waves. Because the mass ratio of the donor and accretor decreases, their orbits expand. Within the age of the Universe they can reach orbital periods of about 80 min \cite[(Deloye \& Bildsten 2003)]{deloye2003}. At longer orbital periods, the mass transfer rates decrease and their orbits expand slower because gravitational wave emission becomes weaker. For this reason, an UCXB is expected to spend most of its life at a long orbital period, and we expect the majority of the population to have periods longer than about 60 min.

Here we address the glaring discrepancy between this theoretical argument, and the fact that all known UCXBs have periods \textit{shorter} than 60 min. At some point in their lives, UCXBs must become either invisible to our instruments, or become disrupted. The latter could be caused by a dynamical instability in the mass loss by the donor. Furthermore, radiation from the accretion disk and millisecond pulsar accretor can potentially evaporate the donor \cite[(Ruderman et al. 1989)]{ruderman1989}. Finally, invisibility may be caused by the magnetosphere of the millisecond pulsar disrupting the inner accretion disk, source of most X-ray emission, and reducing the accretion rate.

\section{Dynamical instability of the donor star}

\begin{figure}[b]
\begin{center}
 \includegraphics[width=3.4in]{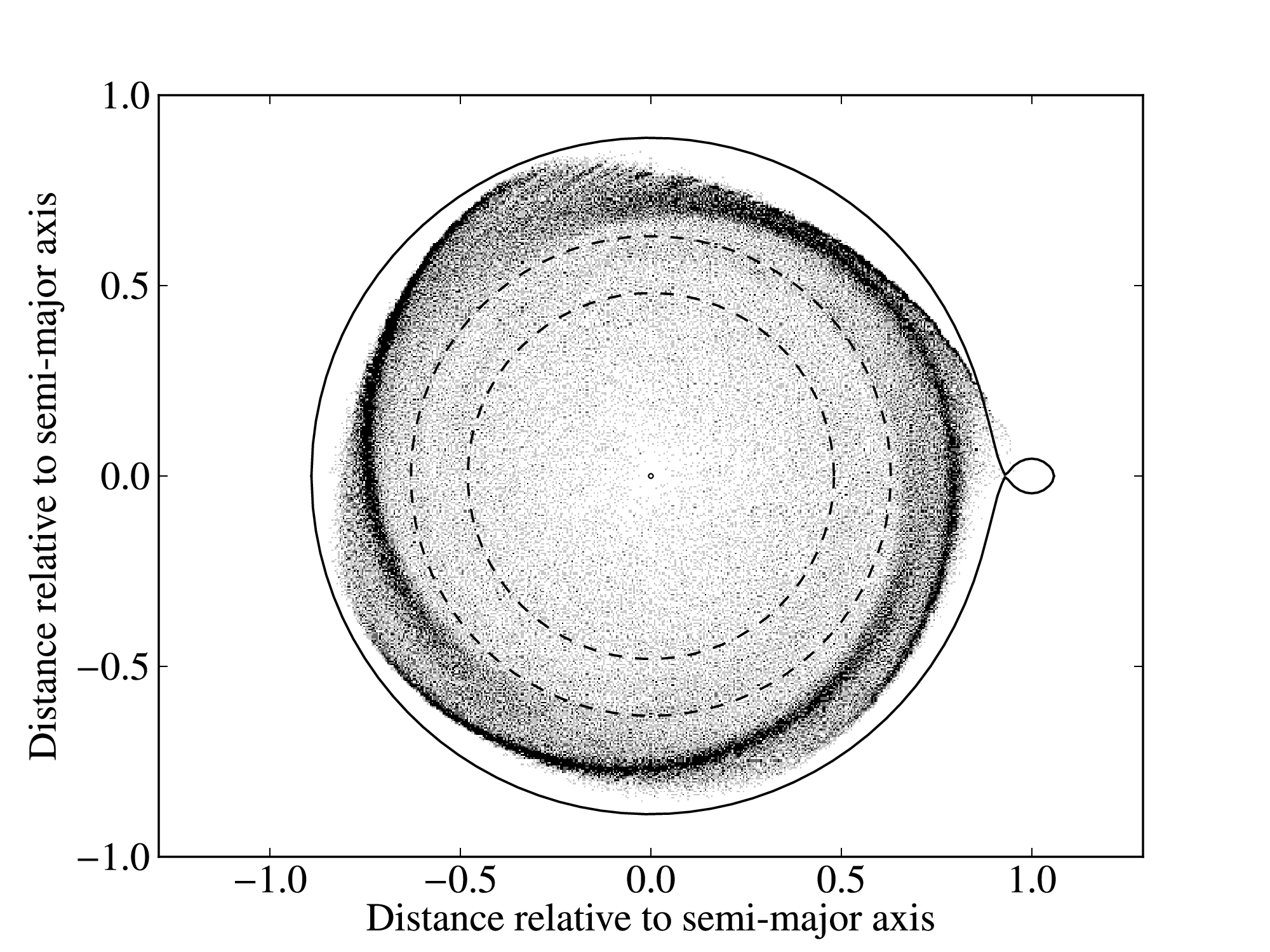}
 \caption{SPH simulation of an accretion disk in dynamical equilibrium for mass ratio of $0.001$. The solid curve shows the equatorial Roche lobes. The dashed circles indicate the 3:1 (inner) and 2:1 (outer) resonances with the orbital period. Figure adapted from \cite{vanhaaften2012evo}.}
   \label{fig:accdisk}
\end{center}
\end{figure}

The evolution of UCXBs and other very low mass ratio binaries is poorly understood. In particular, the dynamical behavior of a large accretion disk (relative to the orbit) is unclear. Along with matter, angular momentum is transported from the donor to the accretion disk. The evolution of an UCXB depends strongly on the capacity of the binary to return angular momentum from the disk to the orbit via a torque. It had been suggested by \cite{yungelson2006} among others that the outwards angular momentum transport could be hampered by gaps in the disk at radii that are resonant with the orbital period, or that the disk would no longer fit inside the Roche lobe of the accretor. At very low mass ratio, reduced feedback has a catastrophic effect on the donor; it cannot be contained in its Roche lobe anymore, with runaway mass loss as a result. By simulating accretion disks in extremely low mass ratio binaries such as the one in Fig.~\ref{fig:accdisk} we find that this feedback mechanism most likely remains effective even at very low mass ratio, and we expect UCXBs to avoid disruption by dynamical instability. This confirmed a result by \cite{priedhorsky1988}.

\section{Evaporation of the donor star}

Last year \cite{bailes2011} discovered a (detached) companion orbiting the millisecond radio pulsar PSR J1719--1438. Remarkable is the extremely low mass function, pointing to a companion mass close to $0.001\ M_{\odot}$, as well as the short orbital period of 131 min. The authors suggested that this system could be the descendant of an UCXB, provided it could become detached at some stage. However, the orbital period of 131 min is significantly longer than the $\sim 80$ min expected by gravitational-wave dominated evolution. Among other things, we investigated the effect of a fast, isotropic wind from the donor on the evolution of an UCXB. Because the donor star in an UCXB contains most of the system's angular momentum in its orbit around the center of mass, this wind is effective in removing angular momentum from the system, besides the losses via gravitational wave emission. The evolution would speed up and longer orbital periods could be reached within the age of the Universe. In \cite{vanhaaften2012j1719} we studied the effect of such a wind on the time it takes for an UCXB to reach a period of 131 min. Even a low wind mass loss rate of a few times $10^{-13}\ M_{\odot}\ \mathrm{yr}^{-1}$ is sufficient, suggesting that UCXBs could indeed produce a system like PSR J1719--1438.

\begin{figure}[b]
\begin{center}
 \includegraphics[width=3.4in]{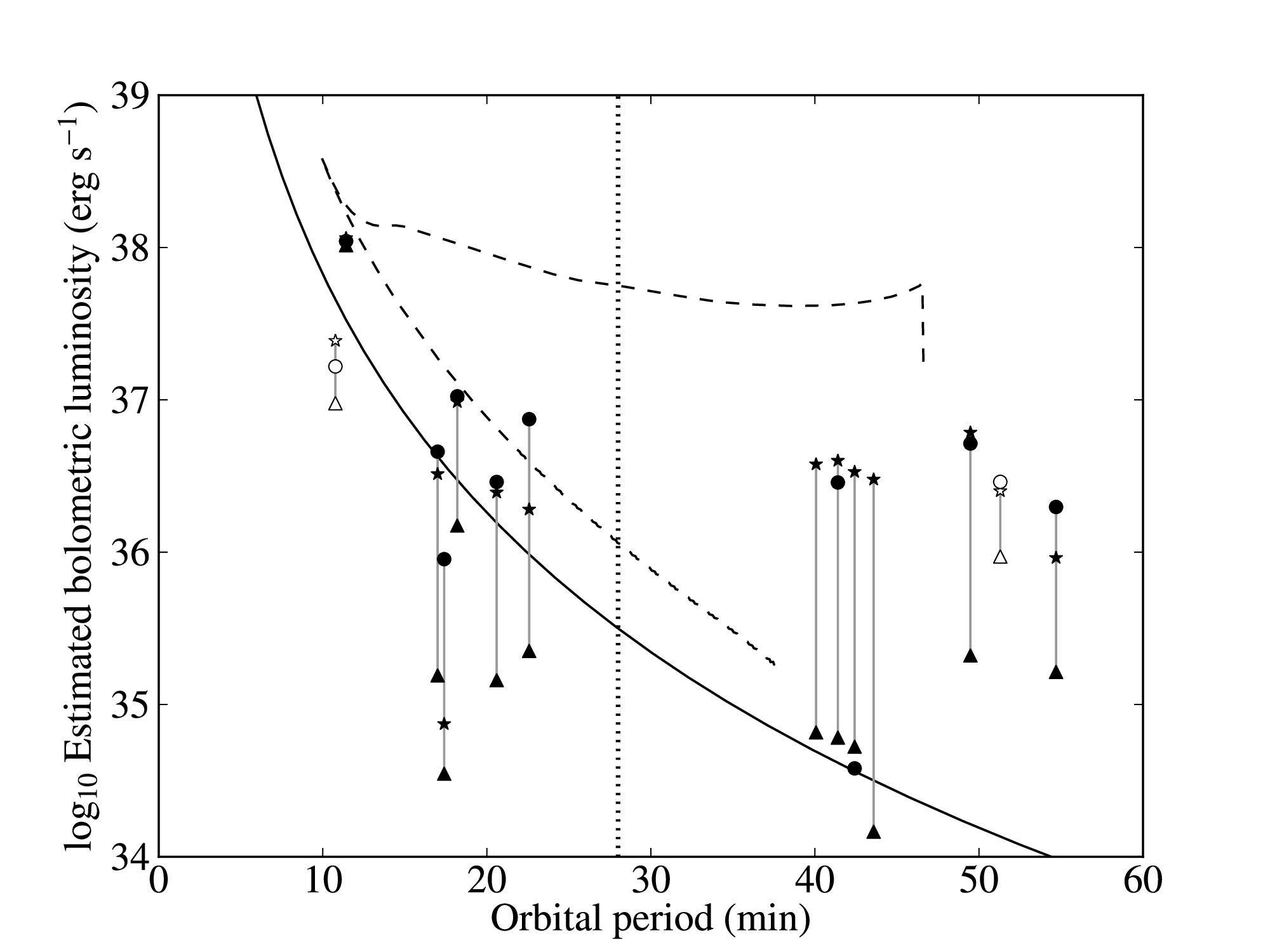}
 \caption{UCXB time-averaged luminosity against orbital period. The solid curve shows the evolution of an UCXB with a helium white dwarf donor. The dashed curve represents the evolution of an UCXBs with initially a helium burning donor \cite[(Nelemans et al. 2010)]{nelemans2010}. The triangles represent the lower bounds on the luminosity from the \textit{Rossi XTE} All-Sky Monitor. The stars above these represent a luminosity estimate based on an extrapolation of the light curves. The circles show the time-averaged luminosities. Filled symbols are confirmed UCXBs, open symbols are candidates. Symbols that correspond to the same source are connected by a gray line for clarity. Figure taken from \cite{vanhaaften2012asm}.}
   \label{fig:asm}
\end{center}
\end{figure}

This hypothesis is supported by 16-year observations by the \textit{Rossi XTE} All-Sky Monitor, summarized in Fig.~\ref{fig:asm}. Most of the UCXBs with orbital periods longer than about 40 min are much brighter than expected from a model with a degenerate donor star and evolution driven by gravitational wave emission. This behavior is consistent with additional angular momentum loss, for example by a donor wind.

Very recently, after this conference, direct evidence of a wind from the donor has been presented with the discovery of PSR J1311--3430 \cite[(Pletsch et al. 2012)]{pletsch2012}, an UCXB with a 93.8 min orbital period \cite[(Romani 2012, Kataoka et al. 2012)]{romani2012,kataoka2012} and a helium donor \cite[(Romani et al. 2012)]{romanietal2012} showing donor evaporation.

\section{The magnetic propeller effect}

\begin{figure}[b]
\begin{center}
 \includegraphics[width=3.4in]{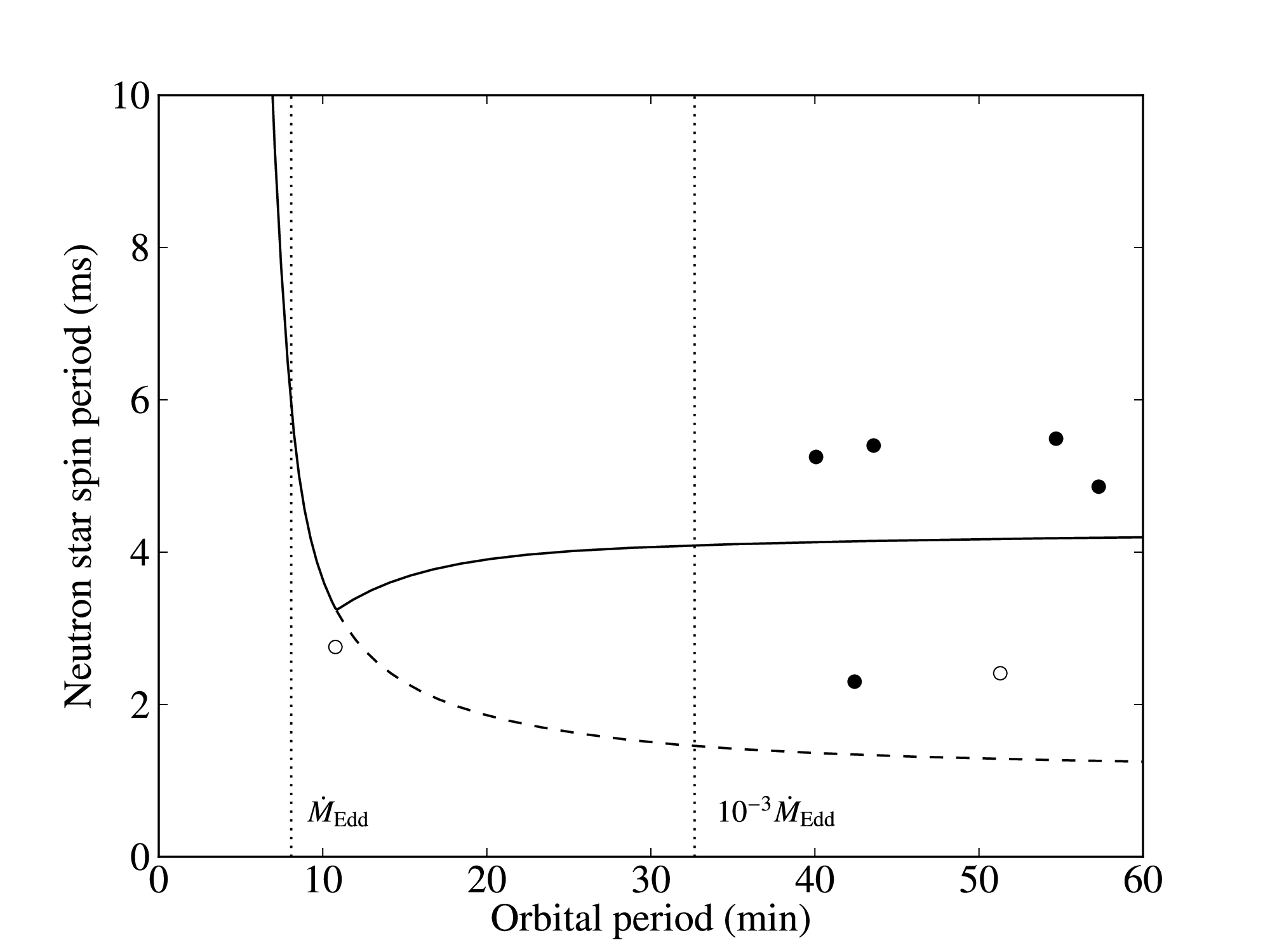}
 \caption{Spin period of a neutron star accretor with a residual magnetic field of $10^{8.5}$ G versus orbital period, with the propeller effect (solid), or in the hypothetical case of continued accreting (dashed). All observed millisecond pulsar-UCXBs are shown by circles (white circles for candidate UCXBs). Figure adapted from \cite{vanhaaften2012evo}.}
   \label{fig:spin}
\end{center}
\end{figure}

If UCXBs survive up to old age and long orbital periods, their mass transfer rates decrease with increasing periods, and the magnetic fields of the neutron stars can more easily dominate the Keplerian flow in the inner accretion disk. For a fast spinning neutron star, matter in the disk can be accelerated by the field lines, leading to a reduction in accretion, or even to matter being ejected from the binary system \cite[(Davidson \& Ostriker 1973)]{davidson1973}. We modeled the spin up and spin down of the neutron star based on the mass transfer rate as function of time in an UCXB. Initially the neutron star spins up due to accretion at a high rate, followed by spin down as the propeller effect becomes effective (Fig.~\ref{fig:spin}). We found that enough rotational energy is stored in the neutron star during the spin-up phase for the propeller effect to prevent accretion completely for the remainder of the evolution. Therefore it is conceivable that the luminosity of an UCXB with a low mass transfer rate is affected by the propeller effect.

\section{Conclusion}

The absence (until very recently) of observed UCXBs with orbital periods longer than one hour could be explained if their evolution is sped up by a donor wind, especially if this eventually leads to detachment of the donor. Furthermore, the propeller effect is capable of significantly reducing the X-ray luminosity of old UCXBs.

\end{document}